# Spectra of $CO_2$-$N_2$ dimer in the 4.2 micron region: symmetry breaking of the intramolecular $CO_2$ bend, the intermolecular bend and higher *K*-values for the fundamental

A.J. Barclay,[a] A.R.W. McKellar,[b] and  N. Moazzen-Ahmadi[a]

[a] *Department of Physics and Astronomy, University of Calgary, 2500 University Drive North West, Calgary, Alberta T2N 1N4, Canada*

[b] *National Research Council of Canada, Ottawa, Ontario K1A 0R6, Canada*

## Abstract

Infrared spectra of the $CO_2$-$N_2$ dimer are observed in the carbon dioxide $v_3$ asymmetric stretch region ($\approx$2350 cm$^{-1}$) using a tunable infrared optical parametric oscillator to probe a pulsed slit jet supersonic expansion. Previous results for the *b*-type fundamental band are extended to higher values of $K_a$. An *a*-type combination band involving the lowest in-plane intermolecular bending mode is observed. This yields a value of 21.4 cm$^{-1}$, and represents the first experimental determination of an intermolecular mode for $CO_2$-$N_2$. This intermolecular frequency is at odds with the value of 45.9 cm$^{-1}$ obtained from a recent 4D intermolecular potential energy surface. In addition, two weak bands near 2337 cm$^{-1}$ are assigned to the $CO_2$ hot band transition $(v_1, v_2^{l2}, v_3) = (01^11) \leftarrow (01^10)$. They yield a value of 2.307 cm$^{-1}$ for the splitting of the degenerate $CO_2$ $v_2$ bend into in-plane and out-of-plane components due to the presence of the nearby $N_2$. The in-plane mode lies at lower energy relative to the out-of-plane mode.

**Introduction**

Interactions between $CO_2$ and $N_2$ molecules are key to measuring and understanding greenhouse effects in the earth's atmosphere since collisions with $N_2$ are the main contributor to pressure broadening of $CO_2$ absorption lines. The essential ingredient which characterizes these interactions is a highly accurate potential energy surface (PES) employed to compute the bound state energies, the dissociation energy, and quantum scattering calculations. The most direct and precise experimental probe of the attractive part of the potential, i.e., probe of bound state energies, are made using rotationally resolved spectra. However, there have been relatively few studies of the weakly-bound $CO_2$-$N_2$ in the gas phase using high resolution techniques. The first of these, published in 1988 by Walsh et al. [1], explored the dimer's infrared spectrum in the region of the $CO_2$ $\nu_3$ fundamental vibration ($\approx$2350 cm$^{-1}$) using a pulsed supersonic jet expansion and a tunable diode laser. They established a T-shaped structure with the N-N axis pointing toward the C atom and an effective intermolecular separation of about 3.73 Å. A similar infrared observation, but using the $^{12}C^{18}O_2$ isotopologue, was made more recently by Konno et al. [2]. Finally, the microwave pure rotational spectrum of $CO_2$-$N_2$ has been investigated by Frohman et al. [3]. Although neither $N_2$ nor $CO_2$ is polar, $CO_2$-$N_2$ has a small permanent dipole moment. The dipole moment lies along the N-N axis and is generated due to the relatively large molecular quadrupole moment of $CO_2$ and to a smaller degree due to the presence of nitrogen molecule causing asymmetric zero-point bending modes in $CO_2$. Aside from the $A$ rotational constant, this study yielded precise rotational and hyperfine splitting parameters for the ground vibrational state.

These three papers represent the only gas phase spectroscopic studies of $CO_2$-$N_2$, but there has also been some infrared work using lower resolution matrix isolation techniques [4,5].

However, these results are more difficult to interpret since interactions with the matrix (usually argon) can be similar in strength to those between the $CO_2$ and $N_2$. Calculations of the structure of $CO_2$-$N_2$ have been reported at various levels of *ab initio* theory [3,5-8], with results most recently including a complete four-dimensional intermolecular potential surface at the CCSD(T)-F12/aug-cc-pVTZ triple zeta level [9] for which the dynamics have been investigated in detail [10].

The present study concerns the infrared spectrum of the normal isotopologue ($^{12}C^{16}O_2$-$^{14}N_2$) in the $CO_2$ $\nu_3$ region, similar to Walsh et al. [1]. However, their results are extended here in three significant directions. First, we observe transitions involving up to $K_a'' = 6$ in the ground state and $K_a' = 7$ in the excited ($CO_2$ $\nu_3$) state. In comparison, only $K_a'' = 0$ was observed in [3], $K_a'' =$ up to 2 and $K_a' =$ up to 3 in [1], and $K_a'' =$ up to 4 and $K_a' =$ up to 3 in [2]. Second, we observe a combination band involving an intermolecular bending vibration, the first such observation for $CO_2$-$N_2$. And third, we resolve and precisely determine the splitting into in-plane and out-of-plane modes of the intramolecular $CO_2$ bending vibration by detecting the $CO_2$-$N_2$ spectrum accompanying the weak $(\nu_1, \nu_2^{l2}, \nu_3) = (01^11) \leftarrow (01^10)$ hot band of $CO_2$.

## 1. Background theory

We note that other isomers of $CO_2$-$N_2$ are predicted theoretically, but here we focus on the observed T-shaped species. This known [1,3,9] equilibrium structure of $CO_2$-$N_2$ is illustrated in Fig. 1. The dimer has $C_{2v}$ point group symmetry. Since the $a$-axis (coinciding with the N-N axis) is perpendicular to the O-C-O axis, rotation around $a$ interchanges the O-atoms, and thus only levels with even $K_a$-values are allowed by nuclear spin statistics in the ground vibrational state, for dimers containing $C^{16}O_2$ or $C^{18}O_2$. The $b$-axis is parallel to the O-C-O axis, so the fundamental band in the $CO_2$ $\nu_3$ (asymmetric stretch) region has *b*-type rotational selection rules. The *c* axis is perpendicular to the plane of the dimer. The four possible intermolecular vibrational modes can be described as: van der Waals stretch ($A_1$ symmetry); out-of-plane rock ($B_1$); and two in-plane bends ($B_2$). These in-plane bends correspond to a $N_2$ or $CO_2$ rock, or else to a geared or anti-geared rotation, depending on their degree of coupling.

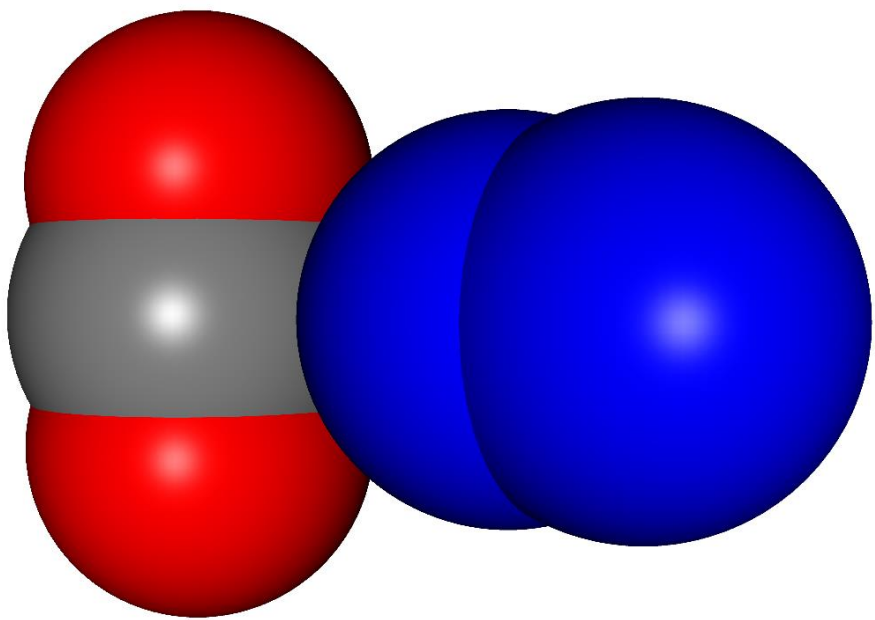

Figure 1: Illustration of the structure of the $CO_2$-$N_2$ dimer. The *a*-inertial axis coincides with the N-N-C axis, the *b*-axis is parallel to the O-C-O axis, and the *c*-axis is perpendicular to the plane of the dimer

 How freely can $N_2$ rotate within the $CO_2$-$N_2$ dimer? In other words, how high in energy is the barrier to interchange of the N nuclei? The fact that no tunneling effects such as doubling were observed in the infrared spectrum [1] suggested that this barrier is fairly high. With much higher resolution in the microwave spectrum, Frohman et al. [3] were able to separately assign transitions corresponding to *para-* and *ortho*-$N_2$, representing the two lowest tunneling components. This effectively shows that interchange is feasible (on the time scale of the experiment). However, the differences between the *para* and *ortho* spectra were very small (i.e.

less than the observed [14]N hyperfine splittings). Transitions between the *para* and *ortho* levels are forbidden, so the actual tunneling splitting cannot be determined experimentally. Calculations indicate that the minimum energy tunneling path involves out of plane $N_2$ rotation, and a simple one-dimensional model based on this path yielded an estimated tunneling splitting of about 6 MHz, or 0.0002 cm$^{-1}$ [3]. A more sophisticated multidimensional calculation [10] has predicted a splitting of less than 0.01 cm$^{-1}$. These splittings apply to the ground vibrational state of the dimer. Larger splittings are expected when intermolecular vibrations, such as the out of plane $N_2$ rock, are excited [10]. In any case, since the small effects due to tunneling are only barely observed in the microwave spectrum, we can be fairly sure that they will remain completely unresolved in the present infrared spectra.

## 2. Observed spectra

Spectra were recorded at the University of Calgary as described previously [11-13]. A pulsed supersonic slit jet expansion is probed using a rapid-scan optical parametric oscillator source. A typical expansion gas mixture contained about 0.04% $CO_2$ plus 1% $N_2$ in helium carrier gas with a backing pressure of about 12 atmospheres. Wavenumber calibration is carried out by simultaneously recording signals from a fixed etalon and a reference gas cell containing room temperature $CO_2$. Spectral simulation and fitting are made using the PGOPHER software [14].

### 2.1. The *b*-type fundamental and *a*-type combination bands

The fundamental was previously studied by Walsh et al. [1]. Our new spectrum, shown in the top panel of Fig. 2, has much better signal-to-noise ratio, covers a wider range, and has somewhat better spectral resolution (our effective line width is about 0.0025 cm$^{-1}$). The wide frequency range enabled us to assign transitions involving six perpendicular subbands ($K_a =$ 3←4, 1←2, 1←0, 3←2, 5←4, 7←6), as compared to three in [1]. We assigned a total of 96 observed lines in terms of 107 transitions. The blended lines fitted to two or more transitions were mostly unresolved asymmetry doublets.

The combination band is shown in the bottom panel of Fig. 2. It is a simple near-prolate parallel (*a*-type) band with $\Delta K_a = 0$, showing relatively strong $K_a = 0$ *P*- and *R*-branches, a prominent $K_a = 2$ *Q*-branch, weaker $K_a = 2$ *P*- and *R*-branches, and very weak $K_a = 4$ *P*-, *Q*-, and *R*-branches. The lower state of this transition is unambiguously the ground state of $CO_2$-$N_2$, as shown by combination differences matching known values from [1], [3], and the present work. We assigned 63 observed lines in terms of 84 transitions.

These two bands were analyzed in a combined fit, resulting in the parameters listed in Table 1. The fit also included the four previously observed [3] $K_a = 0$ microwave transitions (for which we removed the hyperfine splittings, performed a weighted average of the *ortho*- and *para*- species, and gave a relative weight of $10^4$). Blended lines were fitted to an intensity weighted average of their components using the Mergeblends directive of PGOPHER [14]. The root-mean-square deviation of the infrared observations in the fit was 0.00020 cm$^{-1}$, quite similar to our estimated relative experimental uncertainty.

There are no previous results with which to compare the combination band parameters from Table 1. The remaining parameters agree fairly well with previous determinations [1,3], although the differences are often somewhat larger than combined $3\sigma$ error estimates. However, a close look shows very good agreement for key parameters which should be the best determined, namely $(A' - A'')$, $(B' - B'')$ and $(C' - C'')$ for the fundamental infrared band [1], and $(B'' + C'')/2$ from the $K_a = 0$ microwave transitions [3]. (Of course the latter agreement is expected since we incorporated the microwave data in our fit.)

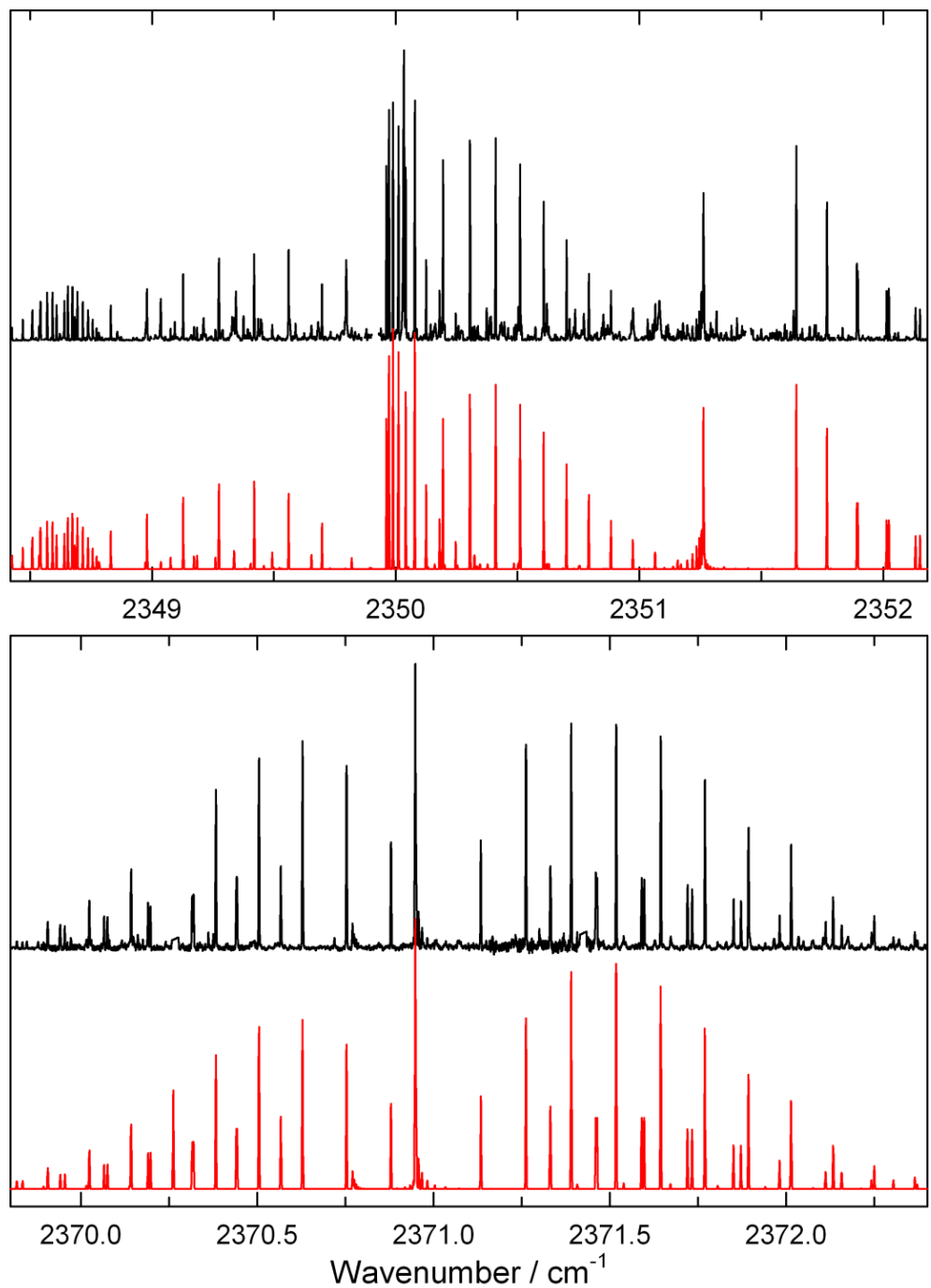

Figure 2: Observed spectrum of CO₂-N₂ in the region of the CO₂ ν₃ fundamental band (top panel) and the combination band involving the lower frequency intermolecular in-plane bending mode (lower panel). Simulated spectra use the parameters in Table 1, with a Gaussian line with of 0.0022 cm⁻¹ and an effective rotational

temperature of 2.2 K. Here and in the other spectrum shown, blank regions correspond to known lines of $CO_2$ monomer and $CO_2$-He dimer.

Table 1. Molecular parameters for $CO_2 - N_2$ (in cm$^{-1}$).[a]

| | Ground state | Excited state | |
| --- | --- | --- | --- |
| | | fundamental | *a*-type combination |
| $\sigma_0$ | 0.0 | 2349.6270(1) | 2371.0063(1) |
| $A$ | 0.396337(19) | 0.393186(12) | 0.381576(29) |
| $B$ | 0.0687903(23) | 0.0687499(27) | 0.0696082(29) |
| $C$ | 0.0581884(23) | 0.0580629(27) | 0.0580675(28) |
| $10^5\,D_K$ | -1.26(10) | -1.387(49) | -2.3(17) |
| $10^5\,D_{JK}$ | 1.593(24) | 1.626(18) | 1.154(24) |
| $10^7\,D_J$ | 3.78(12) | 3.73(25) | 4.03(13) |

[a] Uncertainties (1$\sigma$) in parentheses are in units of the last quoted digit.

## 2.2. The $(01^11) \leftarrow (01^10)$ hot band

We recently detected weak transitions of $CO_2$-Ar [15] in the 2337 cm$^{-1}$ region which could be assigned to the hot band corresponding to $CO_2$ $(v_1, v_2^{l2}, v_3) = (01^11) \leftarrow (01^10)$. This band is observable because a small fraction of dimers in the supersonic expansion are "trapped" with their $CO_2$ constituent in its first excited bending state, $(01^10)$, whose energy is 667 cm$^{-1}$. Trapping occurs because the dominant $CO_2 -$ He collisions in our very dilute gas mixtures are inefficient at relaxing the excited $v_2$ vibration. We estimate this trapped fraction as roughly 1 or 2 %, based on the observed intensity of the hot band compared to the fundamental. Note that the thermal population in the $CO_2$ $(01^10)$ state is about 7.6 % at room temperature, which would presumably be the maximum possible fraction that we could observe in the supersonic expansion from our room temperature nozzle.

The analogous spectrum of $CO_2$-$N_2$ corresponding to the $CO_2$ $(01^11) \leftarrow (01^10)$ hot band is shown in Fig. 3. The $(01^10)$ and $(01^11)$ bending modes are degenerate in the free $CO_2$ molecule. But the degeneracy is broken in the $CO_2$-$N_2$ dimer, splitting the modes into two components, in-plane and out-of-plane, which however are mixed by a Coriolis interaction. This symmetry-breaking effect has not been widely observed in weakly-bound van der Waals complexes, but was detected by Ohshima et al. [16] in $C_2H_2$-Ar, and as mentioned recently by our group in $CO_2$-Ar [15]. The in-plane (i-p) and out-of-plane (o-p) vibrations in the $(01^10)$ lower state have $A_1$ and $B_1$ symmetries, respectively, and those in the $(01^11)$ upper state have $B_2$ and $A_2$ symmetries [15,**Error! Bookmark not defined.**]. The $(01^11) \leftarrow (01^10)$ hot band transitions have $b$-type selection rules, just like the fundamental, with $K_a$ = odd $\leftarrow$ even subbands for the i-p component, and $K_a$ = even $\leftarrow$ odd subbands for the o-p component. There is strong Coriolis mixing between the i-p and o-p modes characterized by a matrix element

$$\langle \text{i-p}, J, k \mid H \mid \text{o-p}, J, k \pm 1 \rangle = \frac{1}{2}\ \xi_b \times [J(J+1) - k(k \pm 1)]^{1/2},$$

where $k$ is signed $K_a$, and $\xi_b$ is the $b$-type Coriolis interaction parameter, related to the dimensionless Coriolis parameter zeta by $\xi_b = 2B\zeta$, where $B$ is the $B$ rotational constant and $\zeta$ can take values between zero (no coupling) and unity (complete coupling).

From the previous studies [15-16], we expect that $\zeta \approx 1$, and also that the $CO_2$-$N_2$ rotational parameters in the $(01^11)$ and $(01^10)$ states should be similar to those in the $(001)$ and $(000)$ states. So the main unknown parameter in assigning the observed spectrum (Fig. 3) was the splitting between the i-p and o-p modes, expected to be similar for $(01^11)$ and $(01^10)$. By trial and error, we found a good simulation of the spectrum for a splitting of about 2.3 cm$^{-1}$, and detailed rotational assignments were then quite straightforward. By far the strongest feature is the very compact and only partially resolved $Q$-branch of the i-p $K_a = 1 \leftarrow 0$ subband at 2337.489 cm$^{-1}$.

The details of our analysis are given in Table 2, and illustrated by the simulated spectra in Fig. 3. We assigned 115 lines (108 transitions) and fitted them with an rms residual of 0.00029 cm$^{-1}$. A few of the transitions are "forbidden" ones connecting the i-p or o-p stacks which become allowed due to the Coriolis mixing. They have $\Delta K_a = 0, \pm 2$, $\Delta K_c = 0, \pm 2$ selection rules, and help to accurately determine the i-p to o-p splitting.

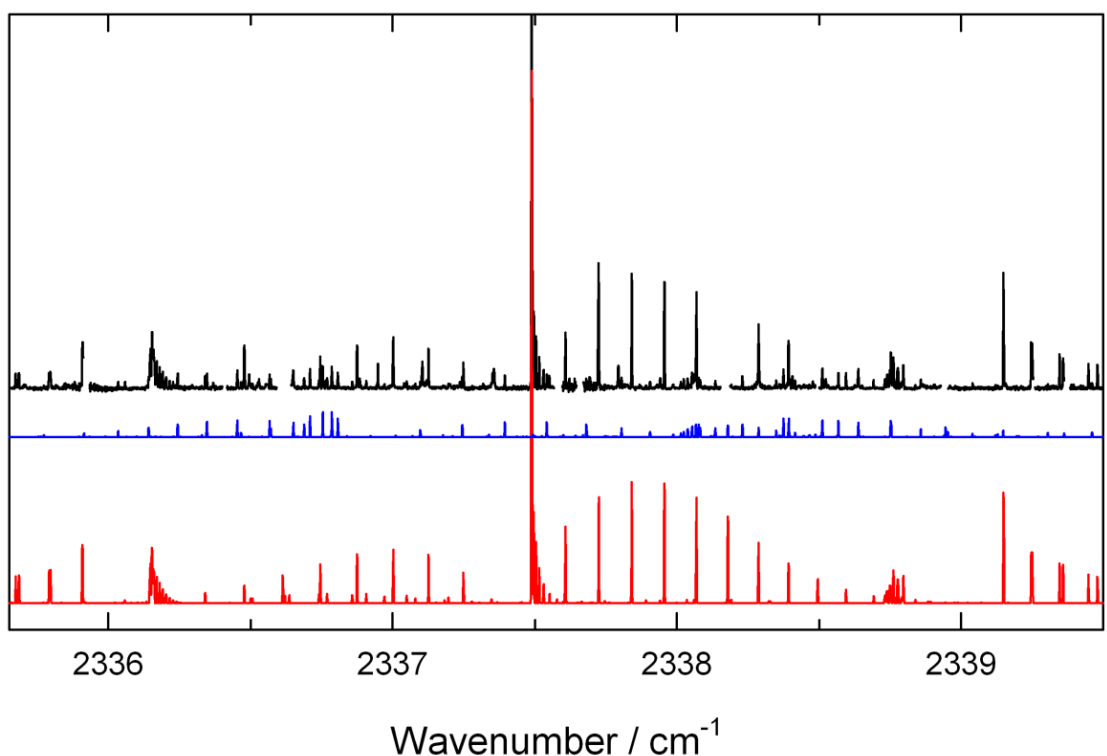

Figure 3: Observed and simulated spectra of $CO_2$-$N_2$ corresponding to the $(v_1, v_2^{l2}, v_3) = (01^11) \leftarrow (01^10)$ hot band of $CO_2$. These bands are affected by Coriolis interaction between the in-plane and out-of-plane components of the $CO_2$ $v_2$ bend. The out-of-plane component (simulated in blue) is weaker since it lies 2.3 cm$^{-1}$ higher in energy than the in-plane component (simulated in red). The very strong feature at 2337.489 cm$^{-1}$ is the mostly unresolved $K_a = 1 \leftarrow 0$ $Q$-branch of the in-plane component.

The lower state of these hot band transitions is the first excited $CO_2$ bending state, ($v_1$, $v_2^{l_2}$, $v_3$) = ($01^10$). This is located at 667.380 cm$^{-1}$ for the free $CO_2$ molecule, but the present spectrum does not tell us exactly what this frequency (called X in Table 2) is for $CO_2$-$N_2$. The shift between the free $CO_2$ value and X is likely to be smaller than a few cm$^{-1}$, and direct observation of the $CO_2$-$N_2$ spectrum in the 667 cm$^{-1}$ region would enable its determination.

Table 2. Molecular parameters for the ($01^11$)-($01^10$) hot band of $CO_2 - N_2$ (in cm$^{-1}$).[a]

|  | ($01^10$) i-p | ($01^10$) o-p | ($01^11$) i-p | ($01^11$) o-p |
|---|---|---|---|---|
| $\sigma_0$ | X | 2.3071(7)+X | 2337.1559(1)+X | 2339.4511(7)+X |
| $A$ | 0.396960(46) | 0.39716(22) | 0.393960(27) | 0.393071(46) |
| $B$ | 0.068690(55) | 0.068975(32) | 0.068458(27) | 0.068901(41) |
| $C$ | 0.058085(20) | 0.058244(14) | 0.0580516(55) | 0.0580988(73) |
| $10^5\ D_{JK}$ | 0.96(19) | 8.7(16) | 0.84(13) | 3.18(21) |
| $\xi_b$ | 0.13832(35) | | 0.13764(19) | |

[a] Uncertainties (1σ) in parentheses are in units of the last quoted digit. $\sigma_0$ is the term value (vibrational energy) of the state. X is equal to the free $CO_2$ $v_2$ frequency (667.380 cm$^{-1}$) plus or minus unknown vibrational shifts which are unlikely to be more than a few cm$^{-1}$. i-p = in-plane; o-p = out-of-plane.

Even though, as just stated, the present spectrum does not give the ($01^10$) ← (000) fundamental origin, it *does* give the exact ($01^11$) ← ($01^10$) hot band origin of $CO_2$-$N_2$, which is 2337.1559 cm$^{-1}$ for the i-p mode. The shift relative to free $CO_2$ is equal to +0.523 cm$^{-1}$ for the i-p mode and +0.511 cm$^{-1}$ for the o-p mode. These values are similar to the shift of the $CO_2$-$N_2$ fundamental band, which is +0.484 cm$^{-1}$.

### 3. Discussion and conclusions

The most interesting result of the hot band analysis is the splitting between the i-p and o-p modes, 2.307 cm$^{-1}$. It may be compared with the values determined for $CO_2$-Ar, 0.877 cm$^{-1}$. In both cases, the o-p mode lies higher in energy. The splitting of the $CO_2$ bending vibration in $CO_2$-$N_2$ has also been detected by argon matrix isolation spectroscopy, with reported values of 2.1 cm$^{-1}$ [4] and 2.0 cm$^{-1}$ [5]. *Ab initio* calculations at the MP2 level have predicted values for this splitting of 3.1 cm$^{-1}$ [5] and 3.82 cm$^{-1}$ [6]. The more recent calculation at CCSD(T)-F12a/aug-cc-pVTZ where all coordinates were fully relaxed gives a splitting of 3 cm$^{-1}$ [9]. All of these calculations were made within the harmonic approximation and in all these cases, the o-p component was above the i-p component, in agreement with our observation.

The values of $\xi_b$ in Table 2 correspond to $\zeta = 1.003$ for the dimensionless Coriolis parameter, similar to the value determined for $CO_2$-Ar. In normal harmonic theory, the maximum value of $\zeta$ is unity, but vibrations in weakly bound complexes tend to be distinctly anharmonic giving rise to anomalous values, $\zeta > 1$.

The difference between the combination and fundamental band origins in Table 1 gives us a value of 21.379 cm$^{-1}$ for a $CO_2$-$N_2$ intermolecular mode. This applies to the dimer with $CO_2$ in the excited $\nu_3$ state, but the value for ground state $CO_2$ is likely to be very similar [12]. The *a*-type character of the combination band indicates that this intermolecular mode has $B_2$ symmetry, and it is almost certainly the lower energy of the two intermolecular in-plane bending fundamentals. The corresponding mode in $CO_2$-CO has values of about 25.5 (C-bonded isomer) and 14.2 cm$^{-1}$ (O-bonded isomer) [**Error! Bookmark not defined.**,17,18]. It is reasonable to see that the $CO_2$-$N_2$ result lies between those of the two isomers of $CO_2$-CO.

Comparison of our 21.379 cm$^{-1}$ intermolecular frequency with previous *ab initio* predictions turns out to be rather tricky. Two published MP2 results give quoted harmonic values of 31.9 cm$^{-1}$ [5] and 3.32 cm$^{-1}$ [6] for the lowest in-plane bend, in rather poor agreement with each other and with our experiment. But the dynamics of $CO_2$-$N_2$ are complex, and so the recent variational calculation [10] based on CCSD(T)-F12 potential surface [9] should be much more reliable, but it gives a value of 45.9 cm$^{-1}$ for this vibration. We suspect that the apparent disagreement with experiment may be due to mislabeling, or failing to locate, intermolecular vibrations, rather than due to errors in the potential surface itself. For comparison, we note the recent *ab initio* calculations [19,20] for $CO_2$-CO, where variational methods predict values for the analogous in-plane bend of 25.5 or 24.45 cm$^{-1}$ (C-bonded isomer) and 14.5 or 14.68 cm$^{-1}$ (O-bonded isomer), in very good agreement with the experimental values as quoted in the previous paragraph. $CO_2$-$N_2$ and $CO_2$-CO have rather similar structures and binding energies, and one might expect that many properties of $CO_2$-$N_2$ might lie between those of the O-bonded and C-bonded isomers of $CO_2$-CO. This indeed appears to be the case, since our observed $CO_2$-$N_2$ frequency (21.379 cm$^{-1}$) lies between the observed $CO_2$-CO values of 14.2 and 25.5 cm$^{-1}$.

Finally, we note the rather large decrease in the value of *A* for the combination band compared to that for the ground state ($\Delta A$ = -0.015 cm$^{-1}$). See Table 1. This is very likely due to an a-type Coriolis interaction between this state and a combination band involving the torsional mode at higher frequency. Such large changes in the values of *A* for the higher energy isomer of $CO_2$-CO have been observed previously [17,18].

In conclusion, spectra of the $CO_2$-$N_2$ van der Waals complex have been studied in the $CO_2$ $\nu_3$ fundamental region ($\approx$2350 cm$^{-1}$) using a tunable infrared OPO laser source to probe a pulsed slit jet supersonic expansion. Previous results for the fundamental band [1] have been extended to higher values of $K_a$, and a combination band involving the lowest in-plane intermolecular mode ($\approx$21.4 cm$^{-1}$) has been observed for the first time. Weak transitions in the 2337 cm$^{-1}$ region have been assigned to the $CO_2$ hot band transition ($v_1$, $v_2^{l2}$, $v_3$) = (01$^1$1) $\leftarrow$ (01$^1$0), and their analysis yielded the splitting of the degenerate $CO_2$ $v_2$ bending vibration into in-plane and out-of-plane modes due to the $N_2$ molecule. This symmetry breaking has a value of 2.307 cm$^{-1}$.

**Acknowledgements**

The financial support of the Natural Sciences and Engineering Research Council of Canada is gratefully acknowledged.